\documentclass[a4paper,fleqn,usenatbib]{mnras}

\usepackage{newtxtext,newtxmath}

\usepackage[T1]{fontenc}
\usepackage{ae,aecompl}

\usepackage{graphicx}	
\usepackage{amsmath}	
\usepackage{amssymb}

\newcommand{\mten}{$\times$\,10$^{10}$\,M$_\odot$}
\newcommand{\kms}{km\,s$^{-1}$}

\title[Stellar dynamics in the Abell 1201 BCG]
{Stellar dynamics in the strong-lensing central galaxy of Abell 1201: 
A low stellar mass-to-light ratio, a large central compact mass, and a standard dark matter halo\thanks{Based 
on observations collected at the European Organisation for Astronomical Research in the Southern Hemisphere under ESO programme 094.B-0823(A).}\thanks{Based on observations made with the NASA/ESA Hubble Space Telescope, obtained from the Data Archive at the Space Telescope Science Institute, 
which is operated by the Association of Universities for Research in Astronomy, Inc., under NASA contract NAS 5-26555. These observations are associated with program 08719.}}
\author[Russell J. Smith et al.]{
Russell J. Smith\thanks{E-mail: russell.smith@durham.ac.uk},
John R. Lucey and Alastair C. Edge
\\
Centre for Extragalactic Astronomy, University of Durham, Durham DH1 3LE, United Kingdom\\
}

\date{Submitted to MNRAS, 13 April 2017; Accepted 20 June 2017.}

\pubyear{2017}

\begin{document}
\label{firstpage}
\pagerange{\pageref{firstpage}--\pageref{lastpage}}
\maketitle

\begin{abstract}
We analyse the stellar kinematics of the $z$\,=\,0.169 brightest cluster galaxy (BCG) in Abell 120, using integral field observations with VLT/MUSE.  This galaxy has a
gravitationally-lensed arc located at unusually small radius ($\sim$5\,kpc), allowing us to constrain the mass distribution using lensing and stellar dynamical information 
over the same radial range.
We measure a velocity dispersion profile which is nearly flat at  $\sigma$\,$\approx$\,285\,\kms\ in the inner $\sim$5\,kpc, and then rises steadily to
$\sigma$\,$\approx$\,360\,\kms\ at $\sim$30\,kpc. 
We analyse the kinematics using axisymmetric Jeans models, finding that the data require both a significant dark matter halo (to fit the rising outer profile)
and a compact central component, with mass $M_{\rm cen}$\,$\approx$\,2.5\,\mten\ (to fit the flat $\sigma$ in the inner regions). 
The latter component could represent a super-massive black hole, in which case it would be among the largest known to date. 
Alternatively $M_{\rm cen}$ could describe excess mass associated with a gradient in the stellar mass-to-light ratio.
Imposing a standard NFW dark matter density profile, we recover a stellar mass-to-light ratio $\Upsilon$ which is consistent with a Milky-Way-like initial mass function (IMF). 
By anchoring the models using the lensing mass constraint, we break the degeneracy between $\Upsilon$ and the inner slope $\gamma$ of the dark matter profile,
finding $\gamma$\,=\,1.0\,$\pm$\,0.1, consistent with the NFW form. 
We show that our results are quite sensitive to the treatment of the central mass in the models. Neglecting $M_{\rm cen}$ biases the results towards both a 
heavier-than-Salpeter IMF {\it and} a shallower-than-NFW dark matter slope ($\gamma$\,$\approx$\,0.5).

\end{abstract}
\begin{keywords}
gravitational lensing: strong -- galaxies: elliptical and lenticular, cD -- galaxies: clusters: individual: Abell 1201 -- galaxies: kinematics and dynamics
\end{keywords}

\section{Introduction}

Cosmological simulations generically predict that, in the absence of baryons, cold dark matter halos have a universal density profile with 
$\rho(r)$\,$\propto$\,$r^{-1}$ in the central regions \citetext{e.g. the ``NFW'' halo of \citealt*{1996ApJ...462..563N}}. Deviations from this profile could result from either 
the influence of non-standard dark matter physics \citetext{e.g. self-interacting particles lead to density distributions with a central core, \citealt{2000PhRvL..84.3760S}},
or a modification of halos by interaction with the baryonic components \citetext{e.g. ``adiabatic contraction'', \citealt{1986ApJ...301...27B}}, or a combination of both. 
Hence obtaining observational limits on the dark matter profile slope can address several important issues relevant to galaxy formation.

For the most massive halos, corresponding to rich galaxy clusters, gravitational lensing and stellar dynamical data have been exploited to attempt to measure
the inner slope of the density profile. In an influential study, \cite{2004ApJ...604...88S} inferred a shallow inner profile $\rho(r)$\,$\propto$\,$r^{-\gamma}$ with
$\gamma$\,$\approx$\,0.5, for three clusters with radial arcs. More recently, \cite{2013ApJ...765...24N,2013ApJ...765...25N} analysed an enlarged sample
of seven clusters, with refinements to the modelling techniques, and reached a similar conclusion, with 
$\langle\gamma\rangle$\,=\,0.50\,$\pm$\,0.17 (including systematic errors).
At face value these results exclude simple NFW halos ($\gamma$\,=\,1), {\it and} ``contracted'' NFW halos (with $\gamma$\,$>$\,1).

One of the key difficulties for this method is that the stellar mass of the brightest cluster galaxy (BCG) itself makes a significant, or even dominant, contribution
to the mass in the innermost few-kpc region of the cluster. The BCG stellar mass-to-light ratio $\Upsilon$ is usually treated as a (constant) free parameter in fitting profile models 
to lensing and dynamical data, 
and the degeneracy between $\Upsilon$ and $\gamma$ is a limiting factor in deriving the latter.
The value of $\Upsilon$ depends sensitively on the stellar initial mass function (IMF), the form of which is not securely established in massive early-type galaxies.
Several studies, using independent methods, have found evidence for an IMF in massive ellipticals which is ``heavier''\footnote{Either ``bottom heavy'' with an 
excess of low-mass stars (relative to solar-mass stars), or ``top-heavy'' with an excess of remnants (relative to solar-mass stars).} (by up to a factor of two)
than that of the Milky Way at given age and metallicity \citetext{e.g. \citealt{2010ApJ...709.1195T}; \citealt{2012ApJ...760...71C}; \citealt{2012Natur.484..485C}}.
However, it is possible that important systematics persist in these analyses \citep{2014MNRAS.443L..69S}.
For example, for some very massive ellipticals, elevated mass-to-light ratios are firmly excluded by strong lensing constraints, despite these same 
galaxies having spectroscopic signatures which can only be fit with bottom-heavy IMFs \citep{2016arXiv161200065N}.
Until such discrepancies are resolved, the stellar population information cannot be confidently used to break the degeneracy with the halo profile. 

\begin{figure*}
\includegraphics[width=180mm]{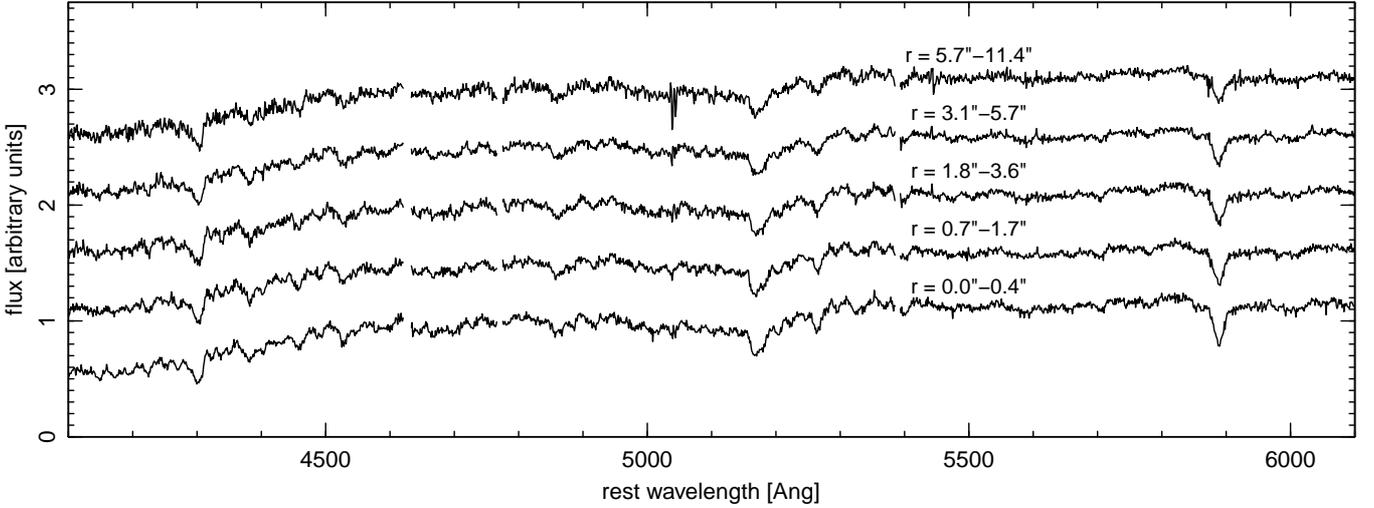}
\vskip -3mm
\caption{
Example spectra from five Voronoi bins (from our preferred binning scheme). Although the bins are spatially distinct, the radial ranges overlap slightly in some cases. Regions affected
by the strongest sky lines have been masked, but residuals from weaker lines are visible in the outermost bin.}
\label{fig:egspec}
\end{figure*}

\begin{figure}
\includegraphics[width=85mm]{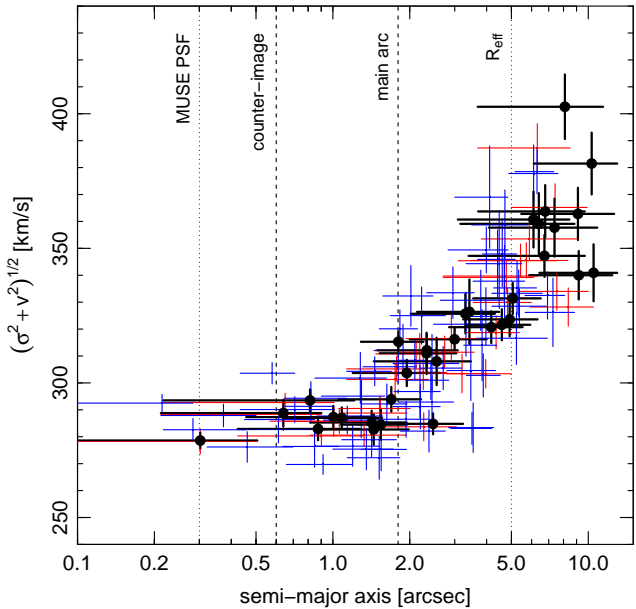}
\vskip -2mm
\caption{The velocity second-moment profile of the Abell 1201 BCG, derived in our preferred Voronoi binning scheme (black points) and for alternative binning choices 
(red and blue). In all cases, the key features of the kinematic data are a flat inner section, and a steeply rising outer profile beyond $\sim$2\,arcsec. Note that the 
velocity dispersion of cluster member galaxies is $\sim$780\,km\,s$^{-1}$ (Owers et al. 2011).}
\label{fig:vorcomp}
\end{figure}

\begin{figure*}
\includegraphics[width=175mm]{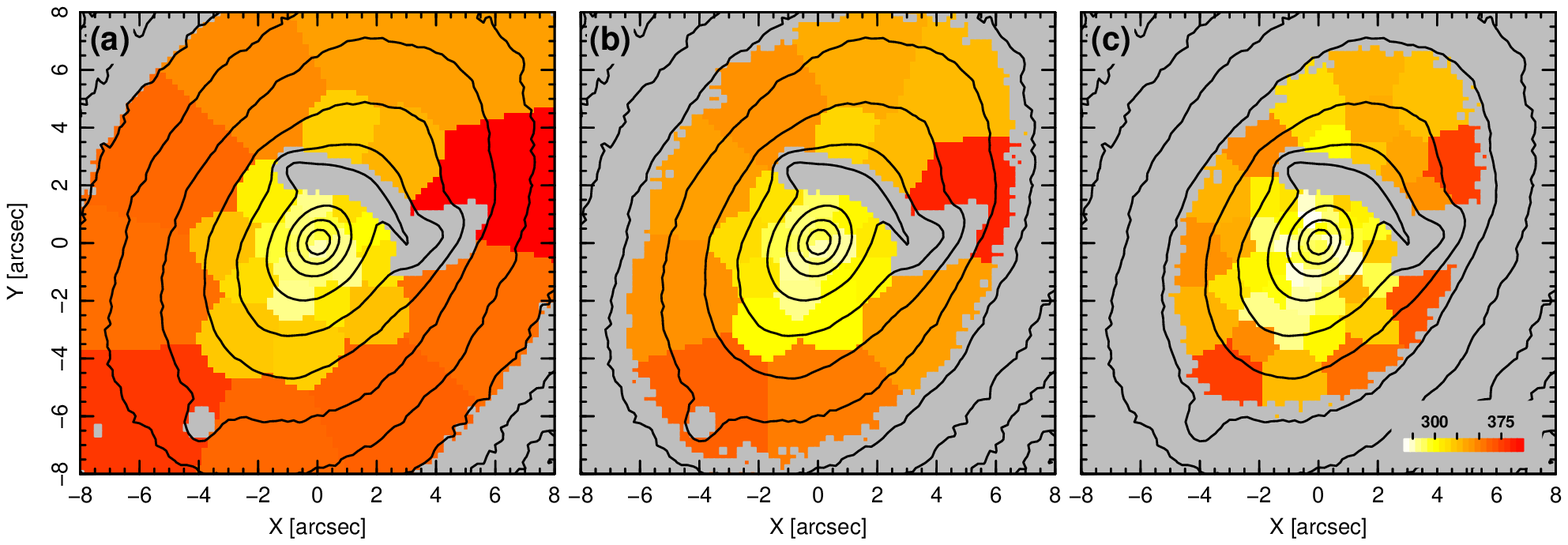}
\vskip -1mm
\caption{Kinematic maps, showing the second velocity moment $(\sigma^2+v^2)^{1/2}$, for the three different Voronoi binning schemes (the same choices  shown in Figure~\ref{fig:vorcomp}).
Systematic velocities are negligible, so this is effectively identical to the $\sigma$ map.
Panel (a) shows the binning scheme used in the dynamical modelling, with $(S/N)_{\rm targ}$\,=\,50 and $(S/N)_{\rm min}$\,=\,1 (black points in Figure~\ref{fig:vorcomp}).
Panel (b) shows the effect of a higher threshold,  $(S/N)_{\rm min}$\,=\,2 (red points in Figure~\ref{fig:vorcomp}) while the scheme in Panel (c) has an even higher threshold and a lower
target-S/N ($(S/N)_{\rm targ}$\,=\,35 and $(S/N)_{\rm min}$\,=\,3), to probe the inner regions with higher resolution (blue points in Figure~\ref{fig:vorcomp}).
In all three cases we recover the radially increasing $\sigma$ trend, broadly aligned with the galaxy isophotes (black contours, from {\it HST} image).}
\label{fig:kinmaps}
\end{figure*}
In the clusters studied by \cite{2013ApJ...765...24N,2013ApJ...765...25N}, the strong lensing constraints on the mass models are usually derived from much larger radius than can be 
probed by the stellar kinematics. Combining information from the different methods in different radial regimes provides leverage over a wide range
of physical scales, but these configurations preclude using the two probes to test assumptions inherent to the methods, or to break the degeneracy
between $\Upsilon$ and $\gamma$, at least for individual clusters (e.g. see figure 1 of \citealt{2013ApJ...765...25N}).
The $z$\,=\,0.169 cluster Abell 1201 offers a rare opportunity to apply this method in an instance where lensing and dynamical constraints {\it overlap} in 
radial scale. Abell 1201 is a post-merger cluster  \citep{2009ApJ...692..702O,2012ApJ...752..139M} with a virial
mass estimated at  $M_{200}$\,=\,(3.9\,$\pm$\,0.1)$\times$$10^{14}$\,M$_{\odot}$ (for $h$\,=\,0.678) from the infall caustic fitting method by \citet{2013ApJ...767...15R}.
From {\it Hubble Space Telescope} imaging, \cite{2003ApJ...599L..69E} discovered an unusual lensed arc around the BCG, at a radius of only 
$\sim$2\,arcsec ($\sim$6\,kpc), which is well within the range amenable to stellar kinematic measurements.

In a previous paper \citetext{\citealt*{2017MNRAS.467..836S}, hereafter Paper I}, we presented new wide-field integral field unit (IFU) 
observations for the BCG of the $z$\,=\,0.169 cluster Abell 1201, focusing on constraints from strong lensing. Plausible mass distributions which reproduce
the  \cite{2003ApJ...599L..69E} arc yield a mass of (34\,$\pm$\,1)\,\mten, where the error reflects the spread among different models. 
We also showed that the presence of a faint inner counter-image to the bright arc requires the total mass profile to be least as steep as the observed luminosity profile. 
We proposed three interpretations of this result. The first possibility (a) is that stellar mass dominates the total profile. This would require a very high 
stellar mass-to-light ratio $\Upsilon$, implying an extremely heavy IMF, and essentially no dark matter within the region probed by lensing\footnote{Assuming the dark matter 
follows a shallower profile than stars on these scales.}.
A second model (b) invoked a steep internal gradient in $\Upsilon$, again probably due to the IMF, but not requiring such extreme deviations 
from the MW mass normalization. The third option (c) was to introduce a central black hole, in which case the required mass was estimated to be  
(1.3\,$\pm$\,0.6)\,\mten, comparable to the largest black holes established from traditional methods.

In this paper, we exploit the same IFU data to measure the stellar kinematics of the Abell 1201 BCG, out to a radius of 30\,kpc, and
place dynamical limits on the relative mass contributions of stars and dark matter. Motivated by the 
additional central mass concentrations envisaged in models (b) and (c) from Paper I, we consider models with an extra dark mass 
at the centre, representing either a super-massive black hole or an enhanced population of dwarf stars or stellar remnants at small radius. We examine
in particular how the introduction of the central mass component affects other inferences from the kinematics, especially with respect to the
the stellar mass-to-light ratio and the dark matter profile slope.

The remainder of the paper follows a simple structure: Section~\ref{sec:kin} describes the data and presents the kinematic measurements, 
while Section~\ref{sec:dyn} describes results from fitting increasingly complex dynamical models. 
Section~\ref{sec:disc} discusses our findings in relation to previous work, and brief conclusions are summarized in Section~\ref{sec:conc}.

As in Paper I, we adopt relevant cosmological parameters from  \cite{2016A&A...594A..13P}:
$h$\,=\,0.678, $\Omega_{\rm M}$\,=\,0.308 and $\Omega_{\rm \Lambda}$\,=\,0.692. In this cosmology, the spatial scale at the 
redshift of Abell 1201 is 2.96\,kpc\,arcsec$^{-1}$.

\section{Kinematic data}\label{sec:kin}

We observed the Abell 1201 BCG using the Multi-Unit Spectroscopic Explorer (MUSE) \citep{2010SPIE.7735E..08B} at the European Southern Observatory 
Very Large Telescope. The observations and data reduction processes were described in Paper I. Briefly, the relevant characteristics are: a uniform exposure time
of 3.1 hours over a 45$\times$45\,arcsec$^2$ central field of view, good image quality (0.6\,arcsec FWHM), and spectral coverage 4750--9350\,\AA, sampled at 
1.25\,\AA\,pixel$^{-1}$ with resolution 2.6\,\AA\ FWHM (at $\lambda$\,=\,7000\,\AA).

We follow the standard approach of defining compact spatial regions of approximately constant total signal-to-noise, using the Voronoi binning method of 
\cite{2003MNRAS.342..345C}. The binning is performed after masking pixels affected by the bright lensed arc, or 
by other background galaxies projected close to the BCG, including the bright spiral at $z$\,=\,0.27. We also exclude all pixels 
below a signal-to-noise threshold of $(S/N)_{\rm min}$\,=\,1 per spatial pixel,
which limits our analysis to a semi-major axis of $\sim$10\,arcsec or $\sim$30\,kpc. The Voronoi bins are defined with a target $S/N$ of 50 (per 1.25\,\AA\ spectral pixel)
at $\sim$5500\,\AA\ in the rest frame. Spectra from some representative spatial bins are shown in Figure~\ref{fig:egspec}.

We measure the mean velocity and velocity dispersion in each Voronoi bin using the penalised pixel fitting method, implemented 
in the {\sc ppxf} software \citep{2004PASP..116..138C}. For templates, 
we allow the code to combine stars of spectral type G5--K5 (giants and dwarfs) from the Indo-US spectral library \citep{2004ApJS..152..251V}, 
which has a native resolution of 1\,\AA\ FWHM.  We fit over a wavelength range of 4150--6150\,\AA\ in the rest-frame, excluding regions around the Mg$b$ triplet and the 
Na\,D lines, which are enhanced in giant elliptical galaxies, and poorly matched by the template stars. We also exclude narrow regions around sky emission lines, 
and regions affected by emission lines from the background galaxies. Errors are derived by Monte Carlo sampling the input spectra.  
(We have tested alternative spectral ranges and masking choices, and found our overall results to be robust against these details.) 

We find that the rotational velocities are negligible, so that the velocity second moment is dominated by the velocity dispersion; for brevity, we use
$\sigma$ as a shorthand for $(\sigma^2+v^2)^{1/2}$, and `velocity dispersion' for `velocity second moment' hereafter.
The measured velocity dispersion profile is shown in Figure~\ref{fig:vorcomp}. The profile is approximately flat from the centre out to $\sim$2\,arcsec, with $\sigma$\,$\approx$\,285\,km\,s$^{-1}$. 
In this region, our measurements overlap with those of  \cite{2004ApJ...604...88S}, who measured a 
significantly smaller velocity dispersion ($\sim$230--250\,km\,s$^{-1}$ within 1.7\,arcsec). 
Similar discrepancies with the Sand et al. measurements, for other clusters, have been reported by
\cite{2013ApJ...765...24N}, who concluded that the earlier data were compromised by poor 
stellar templates and measurement procedures (see discussion in their section 6.4).
Beyond 2\,arcsec, the velocity dispersion increases rapidly to reach $\sim$360\,km\,s$^{-1}$ at $\sim$5--10\,arcsec. 
Rising $\sigma$ profiles of this type appear to be common among BCGs,
as we discuss in Section~\ref{sec:disc}.
In two dimensions, the kinematic profile appears approximately to follow the surface-brightness contours (see Figure~\ref{fig:kinmaps}).
The main features of the kinematic data are robust against changes to our Voronoi binning scheme, e.g. imposing higher $S/N$ threshold and/or 
binning to lower target $S/N$ (Figures~\ref{fig:vorcomp} and \ref{fig:kinmaps}b,c).

\section{Dynamical modelling}\label{sec:dyn}

We analyse the dynamics using the Jeans Axisymmetric Multi-Gaussian Expansion (JAM) method of \cite{2008MNRAS.390...71C}, using 
the {\sc jam} code distributed by the author. This method treats the mass distribution as a collection of oblate ellipsoidal gaussian density 
distributions\footnote{We note that some studies suggest that BCGs are more typically triaxial or prolate \citetext{e.g. \citealt{2010MNRAS.404.1490F}}, 
but Jeans models have been applied to such galaxies in previous works
\citetext{e.g. \citealt{2013ApJ...765...25N}}, and they are a sensible first step before attempting more general but computationally intensive orbit-based methods.}, while the 
velocity dispersion ellipsoid is defined in cylindrical coordinates ($R,\phi,z$), with $\overline{v_R^2}$\,=\,$\overline{v_\phi^2}$ and vertical anisotropy 
$\beta_z$\,=\,$1-\overline{v_z^2}/\overline{v_R^2}$.

We describe the BCG using a multi-component mass model which, in its most general form, incorporates the stellar mass density (derived from the observed luminosity), 
a dark matter halo, and an additional central mass concentration. The following section describes the results of these fits in order of increasing complexity; we summarise the 
general features of the model here. 

The stellar mass component is defined through a Multi-Gaussian Expansion (MGE) \citep*{1992A&A...253..366M} fit to the projected luminosity, 
using the {\sc mge} code of \cite{2002MNRAS.333..400C}. 
For this analysis we use the {\it Hubble Space Telescope (HST)} WFPC2 
(Wide Field and Planetary Camera 2) F606W observations acquired by \cite{2003ApJ...599L..69E}. 
The observed pixel fluxes were calibrated by reference to SDSS $r$-band aperture photometry and corrected to physical units (L$_\odot$\,kpc$^{-1}$), 
adopting a solar absolute magnitude of $M_{r,\odot}$\,=\,4.66 as used in \cite{2005MNRAS.362..799M}. 
A correction of 18 per cent is applied to account for the band-shifting effect, obtained from
the $k$-correction calculator of \cite{2010MNRAS.405.1409C}, using the $g-r$ colour from SDSS.
The translation between luminosity and mass is described by a single stellar-mass-to-light ratio $\Upsilon$ which is quoted in the rest-frame $r$-band.
We assume that the stellar component is viewed edge-on, since the apparent axial ratio is comparable to the most elliptical BCGs \citep{2010MNRAS.404.1490F}.
(Tests for the effects of inclination are summarized in Section~\ref{sec:inctest}).

The dark matter halo is assumed to be spherical for our default models (but see Section~\ref{sec:elliphalo} for the results with elliptical halos), with a density profile following the
generalized NFW (gNFW) form:
\[
\rho(r) \propto (r/r_s)^{-\gamma} (1+r/r_s)^{3-\gamma} \, .
\]
The break radius is fixed at $r_s$\,=\,300\,kpc in all of our models, corresponding to concentration $c$\,=\,5 and $R_{200}$\,$\approx$\,1.5\,Mpc from  \citet{2013ApJ...767...15R}.
The profile break is well beyond the regime probed by the dynamical data, and our observations cannot constrain $r_s$, 
but are in principle sensitive to the asymptotic inner slope $\gamma$, which is unity for the standard NFW halo. 

Additionally, we include an unresolved central mass concentration, treated as an extra gaussian, with very small scale radius, in the MGE.
This component could represent a true super-massive black hole, but could also represent {\it stellar} mass not reflected in the luminosity distribution,  
e.g. due to an increasingly heavy IMF towards the galaxy centre \citep{2015MNRAS.447.1033M,2017ApJ...841...68V}.

We explore the  parameter space  of these dynamical models using 
the {\sc emcee} code by \cite{2013PASP..125..306F}, which implements the ensemble Markov Chain Monte Carlo (MCMC) sampling technique of \cite{2010goodmanweare}.
We use 128 ``walkers'', each making 1000 samples, after a 100-step burn-in. At each step of the MCMC chains, we use {\sc jam} to predict the second velocity 
moment, $(\sigma^2+v^2)^{1/2}$, at the observed pixels, convolved with the seeing of our MUSE data. We assign the pixels to the same Voronoi bins as in the observations,
and  compute the luminosity-weighted average kinematics in each bin. 
Comparing these predictions to the observed values yields the data likelihood for the current parameters of the sampler. 

The lensing models in Paper I provide a robust estimate for the total (dark and luminous) mass projected within an aperture
of radius 4.75\,kpc.  Hence, to facilitate comparison to, and combination with, the lensing information, we describe the model normalisation using 
a projected mass, $M_{\rm ap}$, defined at this radius. 
Most of the models in this paper are independent of the lensing mass. However, 
in some cases (in particular when attempting to constrain the halo profile slope $\gamma$), we incorporate lensing information in a simple way,
by multiplying the likelihood by a prior probability for $M_{\rm ap}$. 
For the prior, we adopt gaussian of mean 34\,\mten, and standard deviation 1\,\mten; this distribution adequately encloses the
estimates derived from all of the plausible models from Paper I.

The results of fitting models within this family are described in Sections~\ref{sec:mfl}--\ref{sec:gnfw}; we explore some model
variations beyond this framework in Section~\ref{sec:addmods}.

\begin{figure*}
\includegraphics[width=175mm]{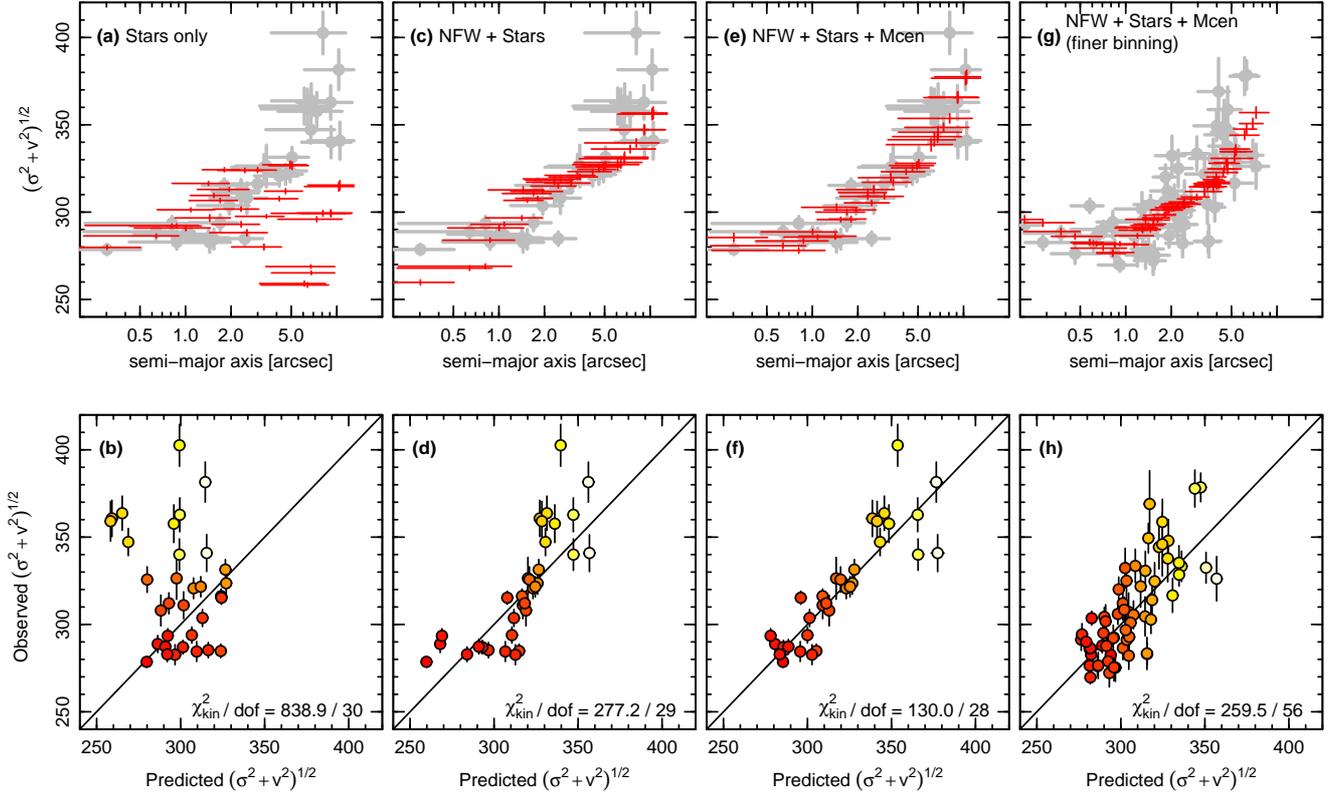}
\vskip -3mm
\caption{Comparison of predictions from selected dynamical models against the observed velocity second moments. The upper panels show the observed kinematic profiles in grey
(same in panels a, c, e), with the predictions from the best-fitting model in each class plotted in red. In the corresponding lower panels, we directly compare observed versus predicted 
values in each spatial bin. The points are colour coded by radius, from inner (red) to outer (white). Panels (g, h) refer to the finer spatial binning scheme shown in Figure~\ref{fig:kinmaps}c.}
\label{fig:kinmodcomp}
\end{figure*}

\begin{table*}
\caption{Best-fitting parameters, and associated quantities, for the models discussed in Sections~\ref{sec:mfl}--\ref{sec:gnfw}
The first five numeric columns list the formal parameters of the fit: 
$M_{\rm ap}$ is the mass projected inside an aperture of 4.75\,kpc, which was determined to be (34\,$\pm$\,1)\,\mten\ from the lensing analysis in Paper I.
$\Upsilon$ is the stellar mass-to-light ratio, in the rest-frame $r$ band. For comparison, a 2\,$Z_\odot$, 9.5\,Gyr-old stellar population ($z_{\rm form}$\,=\,3) has $\Upsilon$\,=\,3.4 
for a Kroupa IMF and $\Upsilon$\,=\,5.3 for a Salpeter IMF.
$M_{\rm cen}$ is the mass of a compact central mass component, which could be interpreted as a black hole.
The anisotropy parameter $\beta_z$\,=\,$1-\overline{v_z^2}/\overline{v_R^2}$ describes the flattening of the velocity dispersion ellipsoid in the ``vertical'' direction.
Finally, $\gamma$ is the asymptotic inner slope of the generalised NFW halo density profile. 
The dark matter fraction $f_{\rm DM}$ is derived from $M_{\rm ap}$, $M_{\rm cen}$ and $\Upsilon$, and refers to the projected fraction inside 4.75\,kpc.
The penultimate column shows the $\chi^2$ accounting only for the 
kinematic predictions, while the final column includes a term penalizing poor predictions of the lensing mass, $\chi_{\rm lens}^2$\,=\,$(M_{\rm ap}-34.0)^2/1.0^2$. This penalty term
is ignored in the fitting, except for the models indicated with $^\star$ in the $M_{\rm ap}$ column.  $M_{\rm ap}$ and $M_{\rm cen}$ are quoted in units of 10$^{10}$\,M$_\odot$.
The headline results of the paper are derived from the models highlighted with bold face.
}
\label{tab:fitpars}
\begin{tabular}{llcccccccc}
\hline
& Model & $M_{\rm ap}$ & $\Upsilon$ & $M_{\rm cen}$ & $\beta_z$ & $\gamma$ &  $f_{\rm DM}$  & $\chi_{\rm kin}^2$/dof & $\chi_{\rm kin}^2+\chi_{\rm lens}^2$ \\
\hline
1 & Stars only & 39.0$\pm$0.4\phantom{$^\star$} & 
& 
						& $-$0.08$\pm$0.02 &  &  & 838.9 / 30 & 863.9 \\
\hline
2& NFW+Stars & 35.8$\pm$0.5\phantom{$^\star$} & \phantom{0}6.7$\pm$0.2 &   & \phantom{+}0.20$\pm$0.02 & [1.0] &  0.36$\pm$0.02 & 277.2 / 29 & 280.4 \\
3 &{\bf NFW+Stars+$M_{\rm cen}$} & 34.6$\pm$0.5\phantom{$^\star$} & \phantom{0}3.9$\pm$0.3 & 2.5$\pm$0.2 & \phantom{+}0.21$\pm$0.02 & [1.0] & 0.54$\pm$0.02 & 130.0 / 28 & 130.4 \\
\hline
4 & gNFW+Stars & 47.1$\pm$1.4\phantom{$^\star$} & \phantom{0}8.1$\pm$0.2 & &  \phantom{+}0.18$\pm$0.02 & 0.07$\pm$0.05 & 0.42$\pm$0.02 & 198.1 / 28 & 369.7 \\ 
5& gNFW+Stars+$M_{\rm cen}$ & 39.2$\pm$5.0\phantom{$^\star$} & \phantom{0}5.9$\pm$1.1 & 1.9$\pm$0.4 & \phantom{+}0.19$\pm$0.03 & 0.58$\pm$0.36 & 0.46$\pm$0.03 &125.7 / 27 & 152.7 \\
6 & {\bf gNFW+Stars+$M_{\rm cen}$} & 34.4$\pm$0.9$^\star$ & \phantom{0}3.9$\pm$0.8 & 2.5$\pm$0.3 & \phantom{+}0.22$\pm$0.02 & 1.01$\pm$0.12 & 0.55$\pm$0.06 & 129.8 / 27 & 130.0 \\
7 &gNFW+Stars & 38.3$\pm$0.7$^\star$ & \phantom{0}7.3$\pm$0.2 & &  \phantom{+}0.23$\pm$0.02 & 0.55$\pm$0.07 & 0.35$\pm$0.01 & 230.0 / 28 &  248.5 \\ 
\hline
\end{tabular}
\end{table*}

\subsection{A mass-follows-light model}\label{sec:mfl}

We begin with the simplest possible model, in which all gravitating mass is described by the MGE fit to the luminosity profile, and
the only free parameters are the projected aperture mass $M_{\rm ap}$, setting the normalisation, and the orbital anisotropy parameter, $\beta_z$. 
The best-fitting solution is reported in line 1 of Table~\ref{tab:fitpars}. The recovered value of $M_{\rm ap}$ translates to a mass-to-light 
ratio of 11.5\,$\pm$\,0.1, which in this case should be interpreted as a {\it total} $M/L$, including any dark matter component as well as stars.
Given the steeply-rising velocity dispersion profile, it is not surprising to find that the mass-follows-light model fails badly to match the observations, in particular 
under-predicting $\sigma$ at large radius (Figure~\ref{fig:kinmodcomp}a,b). The model also yields an aperture mass of (39.0\,$\pm$\,0.4)\,\mten\
which is inconsistent with the lensing constraint at the 5\,$\sigma$ level.

\subsection{Constant-$\Upsilon$ stars and NFW halo}\label{sec:addhalo}

We now introduce an extended dark matter component, assuming a spherical halo that follows the 
NFW density profile. This model is parametrized by $M_{\rm ap}$, $\beta_z$, and 
$\Upsilon$, the latter being explicitly the {\it stellar} mass-to-light ratio.

For a given age and metallicity of the stellar population, the derived value of $\Upsilon$ provides an integral constraint on the IMF, since an excess of low-mass stars (or stellar 
remnants) leads to increased mass, without strongly affecting the luminosity. This constraint is often described through the mass excess factor $\alpha$\,=\,$\Upsilon$/$\Upsilon_{\rm ref}$, 
where $\Upsilon_{\rm ref}$ is the expected stellar mass-to-light ratio given some fiducial IMF. 
For the Abell 1201 BCG, we compute an indicative value of $\Upsilon_{\rm ref}$
from the \cite{2005MNRAS.362..799M} models, adopting an IMF similar to that in the Milky Way. 
For the metallicity, we use the MUSE spectra inside the lensing aperture to measure the [MgFe]$^\prime$ composite index, apply suitable corrections for 
velocity broadening, and compare to the model predictions of \cite*{2011MNRAS.412.2183T}. This method indicates a metallicity of [Z/H]\,=\,0.28\,$\pm$\,0.05.
The observed spectrum has low H$\beta$ absorption, consistent with old stellar ages and there is 
no other evidence for recent or ongoing star-formation in the BCG (no H$\alpha$, no dust features, etc).
Since the spectrum does not constrain the age very tightly, we simply assume formation at early epochs (2.5$<$\,$z_{\rm form}$\,$<$\,4.0)
(corresponding to age 9.5\,$\pm$\,0.5\,Gyr).
Combined with the estimated metallicity and the \cite{2001MNRAS.322..231K} IMF, this yields 
$\Upsilon_{\rm ref}$\,=\,3.4\,$\pm$\,0.2.

The best-fitting parameters for the NFW+Stars model are given in line 2 of Table~\ref{tab:fitpars}. 
The fit attributes a relatively small fraction of the lensing-aperture mass to dark matter 
($f_{\rm DM}$\,=\,0.36), and the stellar mass-to-light ratio is correspondingly high ($\Upsilon$\,=\,6.7\,$\pm$\,0.2). 
Adopting $\Upsilon_{\rm ref}$ from above, this implies $\alpha$\,=\,1.98\,$\pm$\,0.13, indicating substantial deviation from a MW-like IMF.
For comparison, a Salpeter IMF\footnote{A single power-law IMF with the \cite{1955ApJ...121..161S} exponent of 2.35, extrapolated down to 0.1\,M$_\odot$.}
has $\alpha$\,=\,1.55. The derived anisotropy parameter $\beta_z$ is mildly positive, indicating a slight flattening of the velocity dispersion ellipsoid parallel to the symmetry 
axis, which is typical for early-type galaxies \citep{2007MNRAS.379..418C}.

Figure~\ref{fig:kinmodcomp}c,d shows, however, that this model is still a poor fit to the measured kinematics, not only in having a large value 
of $\chi^2$ overall (277 on 29 degrees of freedom), but also in exhibiting clear systematic discrepancies as a function of radius. 
In particular, the central velocity dispersion is under-predicted by $\sim$20\,km\,s$^{-1}$, and $\sigma$ is {\it also} under-predicted in the outermost bins, by $\sim$30\,km\,s$^{-1}$.
There is clearly a tension in this model between matching the steeply rising $\sigma$  profile at large radius, which requires large $f_{\rm DM}$, 
and accounting for the high central $\sigma$, which favours larger stellar contributions.

\subsection{Models with an additional central mass}\label{sec:addmcen}

In Paper I, we showed that the faint inner counter-image to the main arc could be reproduced in simple lensing models only if the 
total mass profile was at least as steep as the observed luminosity profile. As shown in the previous sections, our kinematic data clearly 
require an extended dark matter component, to account for the rising $\sigma$ profile, but the inclusion of dark matter acts to 
{\it flatten} the total mass profile, rather than to steepen it. Hence to reproduce the counter-image, an additional contribution of centrally-concentrated mass 
seems to be required. In Paper I we considered the effects of either an extremely massive central black hole, or additional stellar mass associated with 
a heavier IMF towards the galaxy centre. In either case, the additional mass was (1--4)$\times$10$^{10}$\,M$_\odot$.

Motivated by the lensing results, and by the poor fit of the NFW+Stars model to the kinematics, we now introduce an additional central mass component
into the dynamical fits. We parametrize this as an unresolved mass\footnote{We consider non-point central components in Section~\ref{sec:nonpoint}.}, 
and refer to it as $M_{\rm cen}$, to emphasise that this does not necessarily refer to a {\it true} black hole. 

\begin{figure}
\includegraphics[width=85mm]{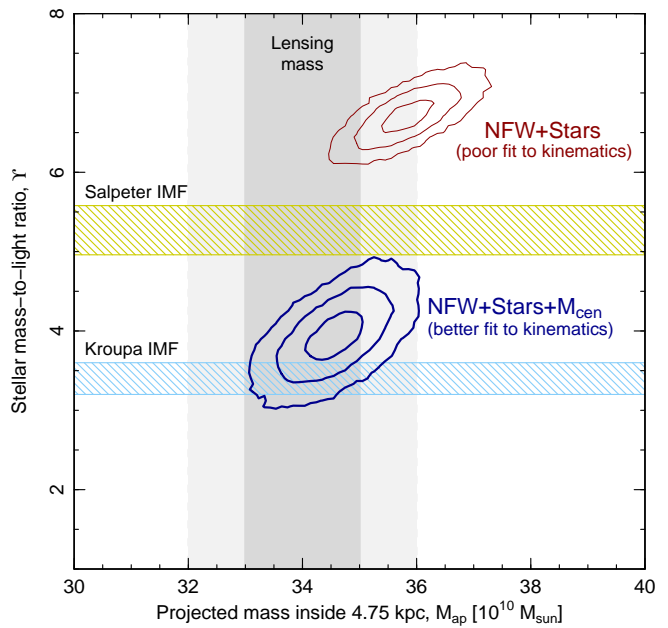}
\vskip -2mm
\caption{Dynamical constraints on the stellar mass-to-light ratio $\Upsilon$, and the arc-enclosed mass $M_{\rm ap}$, for the models with NFW halos. 
The derived parameters for our preferred model (NFW+Stars+$M_{\rm cen}$, which accounts for the presence of a central mass component), 
are shown by the blue contours (1$\sigma$, 2$\sigma$, 3$\sigma$ confidence regions). 
The mass projected inside 4.75\,kpc is consistent with the independent constraint from lensing (grey band), and the stellar mass-to-light ratio is consistent 
with an IMF like that in the Milky Way. The red contours show the parameters recovered if the central mass concentration is neglected in the fit; 
note that this model is a much poorer match to the observed kinematics (Figure~\ref{fig:kinmodcomp}; compare panels e,f versus c,d), and 
yields a substantially larger estimate for $\Upsilon$.}\label{fig:imf_nfw}
\end{figure}

Figure ~\ref{fig:kinmodcomp}e,f confirms that the model with a central mass produces a much better match to the measured velocity
dispersion profile. While the overall $\chi^2$ remains rather large (130 on 28 degrees of freedom), the systematic discrepancies as a function of radius are much reduced.
The introduction of $M_{\rm cen}$ resolves the tension between the solutions preferred at large and small radii, and allows the other two 
components to adjust for improved balance between dark matter and stellar mass. 
Line 3 of Table~\ref{tab:fitpars} reports the parameters of this fit. 
The recovered stellar mass-to-light ratio is now $\Upsilon$\,=\,3.9\,$\pm$\,0.3. 
which is marginally consistent with the value expected for a MW-like IMF, and significantly lighter than expected for a Salpeter IMF. 
(For our adopted $\Upsilon_{\rm ref}$, the mass excess factor is $\alpha$\,=\,1.15\,$\pm$\,0.11.)
Dark matter accounts for a correspondingly larger fraction of the mass inside the fiducial aperture ($f_{\rm DM}$\,=\,0.54).
The recovered central mass 
is $M_{\rm cen}$\,=\,(2.5\,$\pm$\,0.2)\,$\times$\,10$^{10}$\,M$_\odot$,  
consistent with the overall range inferred from lens modelling in Paper I\footnote{But 2$\sigma$ larger than the lensing model which specifically assumes a point-mass, i.e. a true black hole, where
we found $M_{\rm BH}$\,=\,(1.3\,$\pm$\,0.6)\,\mten.}.
The total mass projected inside 4.75\,kpc is $M_{\rm ap}$\,=\,(34.6\,$\pm$\,0.5)\,\mten, in exquisite agreement with the (independent) lensing-derived mass
of $M_{\rm ap}$\,=\,(34\,$\pm$\,1.0)\,\mten.
The joint constraints on $\Upsilon$ and $M_{\rm ap}$ from the models with and without $M_{\rm cen}$ 
are shown in Figure~\ref{fig:imf_nfw}, which highlights the dramatic effect of the central mass on the inferred IMF.

Note that if we treat the central mass as a consequence of gradients in $\Upsilon$, rather than a black hole, 
it is appropriate to include its mass as part of the stellar component when 
quoting the stellar mass-to-light ratio. In this case, the aperture-integrated value (inside the fiducial radius of 4.75\,kpc) is 
$\Upsilon_{\rm ap}$\,=\,$(M_{*,\rm ap}+M_{\rm cen})/L_{\rm ap}$\,=\,$\Upsilon+M_{\rm cen}/L_{\rm ap}$\,$\approx$\,4.7,
which is closer to the Salpeter IMF value, though still much lower than in the NFW+Stars model. 
This result emphasises that the reduction in $\Upsilon$ 
is caused mainly by altering the trade-off between stars and dark matter, rather than directly by $M_{\rm cen}$
itself, which contributes only 7 per cent of the lensing mass. 

\begin{figure}
\includegraphics[width=85mm]{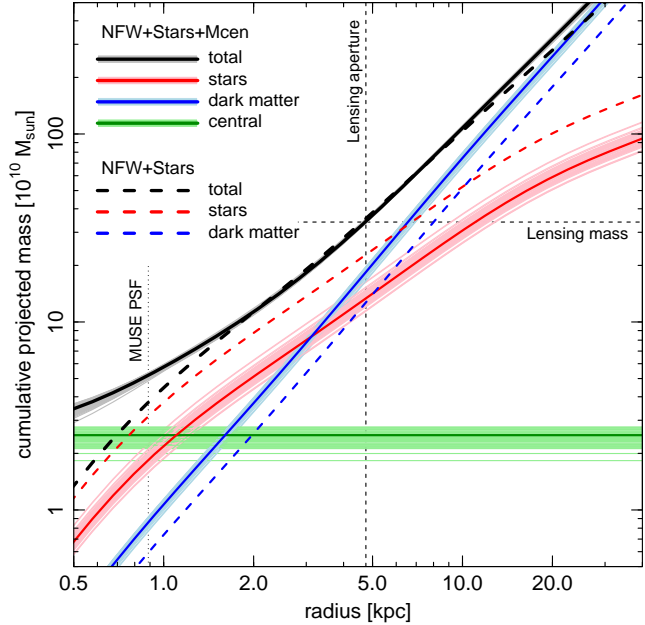}
\vskip -2mm
\caption{Total (solid black) and component (solid colour) projected mass profiles for the model with stars, NFW halo and central mass component. The light bands show samples
from the posterior probability density for the model parameters. The dynamically-derived mass inside 4.75\,kpc is in excellent agreement with the independent constraint from lensing. 
The dashed lines show the best fitting NFW+Stars model, i.e. without a central mass.}
\label{fig:projmass}
\end{figure}

Figure~\ref{fig:projmass} shows the cumulative projected mass profile of our best-fitting model, decomposed into the 
three dynamical components, $M_{\rm cen}$, stars and dark matter. 
Comparing to the equivalent curves for the NFW+Stars model shows that
the {\it total} profiles coincide closely over a decade span in radius (1.5--15\,kpc), but the {\it composition} of the mass differs markedly between the two models.
The NFW+Stars model is dominated by stars within $\sim$10\,kpc, and by dark matter at larger radius. 
In the model with $M_{\rm cen}$, the central component dominates within $\sim$1.2\,kpc, and dark matter 
dominates beyond $\sim$3\,kpc, with stars being the major component in the intermediate range. 

We conclude that when limited to NFW ($\gamma$\,=\,1) dark matter halos, the data require the presence of a central compact mass component, and accounting for this 
mass in the dynamical model has an important impact on the recovery of the other physical parameters, in particular $\Upsilon$.

\subsection{Models with a generalised NFW halo}\label{sec:gnfw}

Finally, we explore models with dark matter following the gNFW profile, with the inner logarithmic density slope $\gamma$ as an additional free parameter. 

In the absence of an additional central mass concentration, the gNFW model reaches a best
match to the kinematic data by flattening the dark matter component completely, and boosting the 
contribution of stars to increase $\sigma$ at small radius. 
The resulting fit (line 4 of Table~\ref{tab:fitpars}) has a high stellar mass-to-light ratio, $\Upsilon$\,$\approx$\,8, 
and a very large projected aperture mass of $M_{\rm ap}$\,$\approx$\,47\,\mten.  Hence although the $\chi^2$ for the kinematics is 
improved compared to the NFW+Stars case, this is achieved at the cost of an unacceptable ($>$10$\,\sigma$) discrepancy with respect to the lensing constraint.

As in the case for NFW halos, including a compact central mass leads to a non-trivial readjustment of the other model components (line 5 of Table~\ref{tab:fitpars}).
The observed kinematics are consistent with a wide range in $\gamma$, though shallower-than-NFW slopes are mildly preferred ($\gamma$\,=\,0.60\,$\pm$\,0.36). 
As before, introducing $M_{\rm cen}$ favours models with a reduced stellar mass and more dark matter within the fiducial aperture.
However, the gNFW model constraints are quite degenerate: shallower halo profiles can be accommodated for higher $\Upsilon$ and lower $M_{\rm cen}$. This 
degeneracy with $\gamma$ translates into increased marginalised errors on the other parameters, e.g. $\Upsilon$\,=\,5.9\,$\pm$\,1.2, which is consistent with any plausible IMF
(e.g. 1.0\,$<$\,$\alpha$\,$<$\,2.5 at 2\,$\sigma$).

\begin{figure}
\includegraphics[width=85mm]{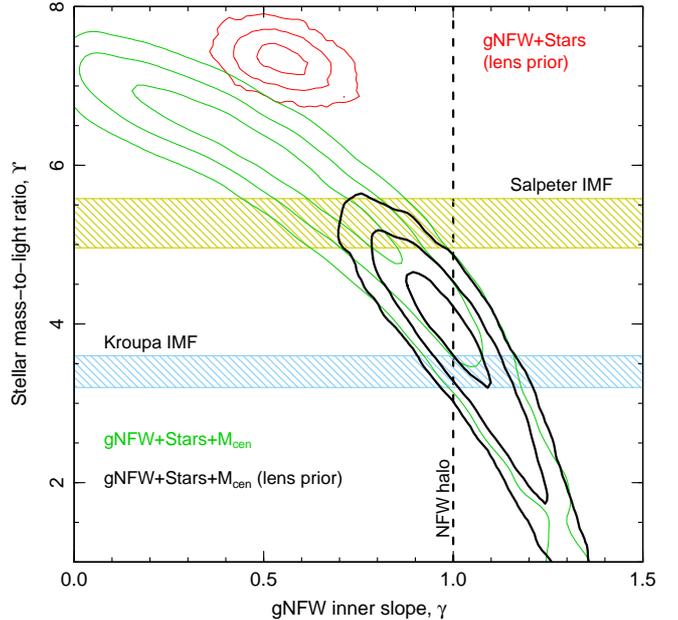}
\vskip -2mm
\caption{Constraints on the stellar mass-to-light ratio $\Upsilon$, and the inner density slope  $\gamma$ of the gNFW profile. Green and black contours show our preferred model, including
a central mass concentration. In the absence of lensing information (green), the fit is very degenerate in this plane. Including the lensing prior favours a halo consistent with the NFW form
(i.e. $\gamma$\,=\,1). The red contours show the disfavoured model in which the central mass is neglected, leading to an overestimate of $\Upsilon$ and an underestimate of $\gamma$.}
\label{fig:gamma_gnfw}
\end{figure}

Along the locus of degenerate models, the low-$\gamma$ solutions imply a larger total mass within the fiducial aperture. 
Hence, we can obtain tighter constraints by explicitly imposing the prior from the lensing configuration, with $M_{\rm ap}$\,=\,(34\,$\pm$\,1)\,\mten. 
The derived parameters for this case are reported in line 6 of Table~\ref{tab:fitpars}. As expected, given the degeneracies with $M_{\rm ap}$, 
the lensing prior favours models with steeper halo profiles, and its inclusion improves the precision on $\gamma$ (see Figure~\ref{fig:gamma_gnfw}). 
The combined lensing and dynamical constraint is consistent with the standard NFW
form: $\gamma$\,=\,1.01\,$\pm$\,0.12. Using the lensing prior also improves the limits on the 
stellar mass-to-light ratio, to $\Upsilon$\,=\,3.9\,$\pm$\,0.8, but this remains compatible with either a Kroupa or Salpeter IMF
($\alpha$\,=\,1.15\,$\pm$\,0.25).
As before, attributing all of the excess mass to the stars, instead of to a black hole, would increase this estimate (integrated within the fiducial aperture) by $\sim$20 per cent.
Because the gNFW+Stars+$M_{\rm cen}$ model returns $\gamma$\,$\approx$\,1, when including the lensing prior, the predicted kinematics are effectively identical 
to those of the NFW+Stars+$M_{\rm cen}$ case shown in Figure~\ref{fig:kinmodcomp}e,f.

For comparison, we also considered a gNFW-halo model incorporating the lensing prior but {\it without} the central mass component. In this case, there is clear tension between the prior and the dynamical 
information, resulting in a best fit which still exceeds the lensing constraint by $\sim$4\,$\sigma$, while the fit to the kinematics is also poor, similar to the equivalent NFW fit without $M_{\rm cen}$. 
The derived mass-to-light ratio is 7.3 (somewhat 
above expectations for a Salpeter IMF), while the preferred halo slope is much flatter than NFW, $\gamma$\,=\,0.55\,$\pm$\,0.07 (Table~\ref{tab:fitpars}, line 7). 

We conclude that in the Abell 1201 BCG, the need for any deviation from an NFW profile is strongly affected by the inclusion or otherwise of 
a central mass component. When the central mass is neglected, we recover a solution with a heavy IMF and a flattened halo profile, similar to the results
of \cite{2013ApJ...765...25N} for more distant and more massive clusters. By contrast, when allowing for the presence of a compact component, comparable in mass to the 
largest-known central black holes, our results favour an orthodox dark matter halo, and are {\it also} consistent with a standard, MW-like IMF. 

\subsection{Additional models and robustness tests}\label{sec:addmods}

In this section we briefly explore some variants on the fitting method, testing for robustness to some of the assumptions made above.

\subsubsection{Non-edge-on inclination?}\label{sec:inctest}

The default models assumed an edge-on configuration, $i$\,=\,90$^\circ$ (i.e. the shortest axis of the oblate ellipsoids are perpendicular to the line of sight),
but the MGE description for the stellar mass is consistent with inclinations as low as $i$\,=\,68$^\circ$. (The limit arises from the highest-ellipticity gaussian in 
the fit to the projected luminosity.) Allowing inclination as a free parameter we find (a) that high inclinations ($i$\,$>$\,80$^\circ$) are somewhat favoured, and 
(b) that none of the parameters of interest to us has significant covariance with $i$. 
Hence, no information is lost by neglecting inclination as a parameter in the models discussed so far.

\subsubsection{Non-point mass central concentrations?}\label{sec:nonpoint}

In our default fits, the central mass component is modelled using a gaussian with very small radius,  $R_{\rm cen}$\,=\,0.01\,arcsec (standard deviation).
While this may be appropriate if $M_{\rm cen}$ represents a true black hole, if the extra mass is instead due to stellar population gradients, a more extended
distribution might be more suitable. 

We have considered alternative models in which $R_{\rm cen}$ is allowed to vary up to a maximum of 2\,arcsec ($\sim$6\,kpc) (but this component is always spherical.) 
The additional freedom in this case leads to stronger degeneracies between $\Upsilon$ and  $M_{\rm cen}$, since the model can trade mass between the 
stars and the extended ``extra'' component, which can now have more similar spatial scales. 

Formally, point-mass contributions are disfavoured by the fits, but the recovered sizes remain very small, and 
comparable to the resolution of the kinematic data: $R_{\rm cen}$\,=\,0.22\,$\pm$\,0.05\,arcsec (i.e. $\sim$0.5\,arcsec FWHM).
The mass attributed to the extra component is much larger than in the point-mass case, but the component following the light is correspondingly reduced. 
Adding both components to describe the total stellar mass yields an aperture-integrated stellar mass-to-light ratio of
$\Upsilon_{\rm ap}$\,=\,4.5\,$\pm$\,0.3, which is similar to the equivalent result for an unresolved central mass. 

\subsubsection{Non-spherical halo?}\label{sec:elliphalo}

X-ray observations have shown that Abell 1201 has a complex morphology, with high ellipticity, cold fronts and an offset core.
The overall configuration has been interpreted as indicating a late-stage merger aligned with the BCG major axis  \citep{2009ApJ...692..702O,2012ApJ...752..139M}.
It is unclear whether such features on $>$100\,kpc scales should invalidate the assumption of a spherical halo on the $\la$20\,kpc scales probed by our data, but motivated by the
X-ray information, we have experimented with including halo ellipticity as a free parameter in the dynamical model. 

Formally, we find that the data do prefer very elliptical models, where the halo has an axial ratio of $q$\,$\approx$\,0.4. In this case, we also find $\beta_z$\,$\approx$\,0, with
a very high dark matter fraction ($\sim$70 per cent) and low stellar mass-to-light ratio $\Upsilon$\,$\approx$\,3 (or $\Upsilon$\,$\approx$\,4 if the central component is 
added to the stellar mass). Crucially, however, the elliptical halo model also yields a very high total mass within the fiducial aperture, $M_{\rm ap}$\,$\approx$\,45\,\mten, 
which is incompatible with the lensing analysis. 
In this model, the elliptical halo can trade off much more readily against the stellar mass distribution, 
and so the three components are shuffled to make small improvements in the fit to the kinematics, at the expense of severely over-predicting the lensing mass. 
Imposing the lensing prior (as for the gNFW model) to suppress these unacceptable solutions, we obtain best-fitting parameters that are similar to the spherical halo 
case ($\Upsilon$\,=\,3.5, $M_{\rm cen}$\,=\,2.7\,\mten), and only a modestly flattened halo shape, $q$\,$\approx$\,0.8.

\subsubsection{Resolution test}

Finally, as a test for robustness against the Voronoi binning assumptions, we have run our preferred NFW+stars+$M_{\rm cen}$ model using the
alternative spatial binning scheme shown in Figure~\ref{fig:kinmaps}c. 
Because a higher S/N threshold was used, the data in this case do not extend to the outermost pixels used in the default fits, 
but they do provide improved sampling in the inner regions.

The fit results for this case are broadly consistent with the default binning scheme, with a slightly increased central mass $M_{\rm cen}$\,=\,(2.9\,$\pm$\,0.2)\,\mten, 
and slightly increased normalization $M_{\rm ap}$\,=\,(35.7\,$\pm$\,0.5)\,\mten). The stellar mass-to-light ratio remains consistent with MW-like IMFs, $\Upsilon$\,=\,4.1\,$\pm$\,0.3
($\alpha$\,=\,1.20\,$\pm$\,0.12). 
The measured and predicted kinematics for this fit are shown in Figure~\ref{fig:kinmodcomp}g,h. As a result of the increased resolution in the inner bins, 
the central mass now generates a noticeable rise in $\sigma$ at the smallest radii, rather than simply flattening the profile as in  Figure~\ref{fig:kinmodcomp}a--f. This signature is seen
in both the data and the model predictions.

\section{Discussion}\label{sec:disc}

\begin{figure}
\includegraphics[width=85mm]{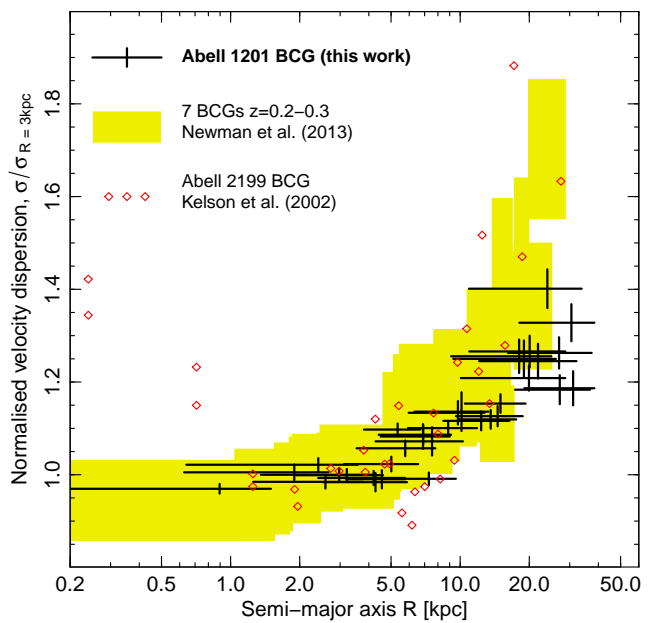}
\vskip -2mm
\caption{Comparison of the velocity dispersion profile to previous measurements for BCGs.
Note that the normalisation is made at 3\,kpc, following Newman et al. (2013). The rise in the inner regions of Abell 2199 (Kelson et al. 2002) 
may be due to resolving the mass contributions from the central black hole in this nearby ($z$\,=\,0.03) BCG.}
\label{fig:newmancomp}
\end{figure}

Our MUSE data have revealed a rising kinematic profile in the Abell 1201 BCG, with velocity dispersion increasing from 285\,km\,s$^{-1}$ inside 5\,kpc to 360\,km\,s$^{-1}$ at $\sim$20\,kpc. 
Such trends are not seen in nearby, lower-luminosity, early-type galaxies, where flat or falling $\sigma$ profiles predominate \citetext{e.g. \citealt{2017A&A...597A..48F}}. 
In BCGs, by contrast, many previous works have 
found evidence for rising profiles in individual galaxies \citetext{e.g. \citealt{1979ApJ...231..659D}, \citealt*{1999MNRAS.307..131C}, \citealt{2002ApJ...576..720K}, \citealt{2011A&A...531A.119R}}.
More recently, \cite{2013ApJ...765...24N} found rising $\sigma$ profiles in all seven BCGs in their sample at $z$\,=\,0.2--0.3, 
with an apparent homogeneity among the profiles after normalization at 3\,kpc. 
\cite{2017MNRAS.464..356V} also report rising $\sigma$ trends for the most massive local ellipticals in their sample (most of which are cluster- or group-dominant galaxies), while 
less massive ellipticals show flat or falling profiles.
Hence at an empirical level, the association of rising profiles with cluster-scale dark matter halos seems to be
well established. Figure~\ref{fig:newmancomp} shows that our measurements in Abell 1201 are  similar to the profiles of the \cite{2013ApJ...765...24N} BCGs, though 
the upturn at large radius is slightly less pronounced. The difference qualitatively accords with Abell 1201 being less massive than average for the Newman et al. clusters. 
As argued in Paper I, comparable contributions of dark and stellar material projected inside 4.75\,kpc are compatible with dark matter halos extracted from 
cosmological simulations  \citep{2007MNRAS.381.1450N,2015MNRAS.446..521S} to mimic the properties of Abell 1201. 
Hence the observed $\sigma$ profile can be generated with a halo mass which is quite plausible in the cosmological context.

Of the three explanations proposed in Paper I to account for the lensing counter-image, one requires the lensing mass to be fully dominated by the
stars\footnote{Or by mass distributed just as steeply as the stellar component, unlike any plausible dark matter candidate.}, with a very 
high stellar mass-to-light ratio.
The rising velocity dispersion profile measured here is inconsistent with this scenario, and 
translates directly into a need for substantial mass contributions from an extended dark matter halo. 
However, the resulting combined profile of dark and stellar mass is too shallow, in the inner 5\,kpc, to reproduce the lensing configuration, in particular the inner counter-image.
Hence with the ``all-stellar'' option ruled out on dynamical grounds, the lensing configuration {\it alone} already implies the presence of additional centrally-concentrated mass,
in the form of either a black hole or a gradient in the stellar mass-to-light ratio.

Independently of the lensing analysis, our kinematic measurements reinforce the need for a compact central mass which is 
not reflected in the luminosity distribution in Abell 1201. The dynamically-derived $M_{\rm cen}$\,=\,(2.5\,$\pm$\,0.2)\,\mten\ is $\sim$2$\sigma$ larger by than the lensing-inferred 
black-hole mass of (1.3\,$\pm$\,0.6)\,\mten. If $M_{\rm cen}$ is really a black hole, its mass would be slightly larger than any in the \cite{2016ApJ...831..134V} compilation, 
as well as an order of magnitude higher than the average $M_{\rm BH}$ at the velocity dispersion of the Abell 1201 BCG. 
Over-massive black-holes in BCGs, compared to the standard scaling relations, 
have been proposed as a means to reconcile radio and X-ray properties of cluster cores \citep{2012MNRAS.424..224H}, and {\it some} local BCGs indeed 
have dynamical $M_{\rm BH}$ which exceed the relations by factors of 4--10 \citep{2011Natur.480..215M}. 
Our result for Abell 1201 could thus be interpreted as simply a more extreme example of a trend already hinted at in nearby, generally less massive, clusters.

If the Abell 1201 BCG really harbours such a large black hole, its low-level accretion activity might generate detectable radio and/or X-ray emission, 
as quantified in the ``Black Hole Fundamental Plane'' \citep*{2003MNRAS.345.1057M}. 
Unfortunately only upper limits are available for the core of Abell 1201. 
The best current radio limit at 5\,GHz is from \citet{2015MNRAS.453.1201H}, who quote a 3$\sigma$ upper limit 
which translates to a radio power of $<$\,1.4$\times$10$^{38}$\,erg\,s$^{-1}$. 
For the X-ray constraint, we estimate a limit on emission coincident with the core of the BCG, from two separate 40\,ks {\it Chandra} observations, 
following \citet{2011MNRAS.413..313H}. Assuming an unobscured power law with an index of --1.7, we derive a 3\,$\sigma$ upper limit to the
unabsorbed 0.2--10\,keV X-ray luminosity of 
2.1$\times$10$^{40}$\,erg\,s$^{-1}$.
Placing these limits on the Fundamental Plane, using the 
scaling relations from \citet{2003MNRAS.345.1057M} or \citet{2012MNRAS.419..267P},
shows that Abell 1201 lies at the lower boundary of the distribution of BCGs with a joint radio and X-ray detection from the sample of \cite{2012MNRAS.424..224H}.
Therefore neither upper limit is inconsistent with the presence of such a high-mass black hole. 
Future Very Long Baseline Array observations could in principle improve on the current radio limit, e.g. by a factor of $\sim$15 for a full track, reaching a radio power of
10$^{37}$\,erg\,s$^{-1}$. However, the X-ray luminosity predicted from the Fundamental Plane 
is $<$\,10$^{38}$\,erg\,s$^{-1}$, comparable to the brightest accreting binaries in the galaxy \citep{2004ApJ...611..846K}, which precludes
obtaining an unambiguous matching detection.

If the central mass is {\it not} a black hole, the most natural explanation is that this component reflects an increased stellar mass-to-light ratio towards the BCG centre. 
Modest radial variations in $\Upsilon$ could be caused by metallicity or age gradients, but as noted in Paper I, the spectra do not indicate
strong gradients in the inner 2\,arcsec. Hence a radial variation in the IMF is probably needed to generate the steep trend in $\Upsilon$ implied by the lensing models.
Some analyses of spectroscopic gradients in nearby ellipticals do support trends in the IMF within half the effective radius (0.5\,$R_{\rm eff}$),
\citetext{e.g. \citealt{2015MNRAS.447.1033M}, \citealt{2016MNRAS.457.1468L}, \citealt{2017ApJ...841...68V}}, but others have argued that the spectral signatures
are consistent with metallicity gradients \citep*{2016ApJ...821...39M,2017MNRAS.468.1594A}. For Abell 1201, the best-studied IMF-sensitive features are redshifted 
beyond the spectral range of our MUSE data, so we cannot currently address this question directly for this galaxy. 
As described in Section~\ref{sec:nonpoint}, our kinematic data seem to favour a compact central mass, in preference to a spatially extended component.
At face value, this disfavours the IMF gradient solution.  For example, figure 17 of \cite{2017ApJ...841...68V} indicates an enhancement of the (aperture-integrated) mass-excess
factor at a characteristic scale of $\sim$0.3\,$R_{\rm eff}$, corresponding to $\sim$5\,kpc in the Abell 1201 BCG. This is an order of magnitude larger than 
the preferred scale recovered when we fit for $R_{\rm cen}$ as a free parameter.

\begin{figure}
\includegraphics[width=85mm]{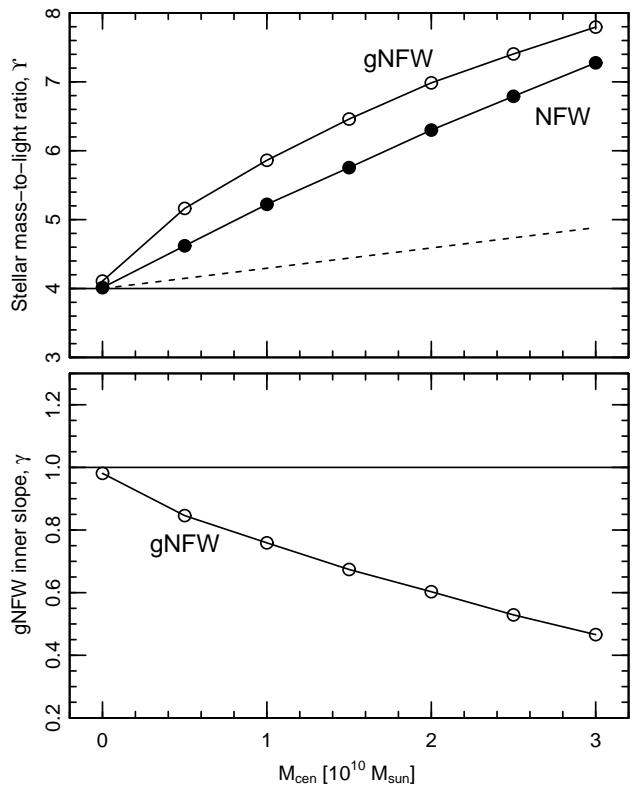}
\vskip -2mm
\caption{Bias in recovery of stellar mass to light ratio $\Upsilon$ and gNFW inner slope $\gamma$, when neglecting the presence of a central point mass $M_{\rm cen}$. 
These estimates are based on idealised realisations of the Abell 1201 BCG modelling. Specifically, we generate mock kinematic data from models with $M_{\rm cen}$\,$>$\,0 (as well
as a stellar component derived by deprojecting the observed luminosity assuming $\Upsilon$\,=\,4.0, and an NFW dark matter halo), and fit with models which force 
$M_{\rm cen}$\,$=$\,0. The dashed line in the upper panel shows the mass-to-light ratio derived within the 4.75\,kpc lensing aperture
if the central mass component is added to the stellar mass.} 
\label{fig:bhmatters}
\end{figure}

Regardless of whether the central mass is actually a black hole or not, we have shown that its treatment in the dynamical model has important implications 
for other parameters determined from the stellar kinematics. In particular, including $M_{\rm cen}$ leads to a significant reduction in the stellar mass-to-light ratio, 
to values compatible with a Kroupa (Milky-Way-like) IMF, rather than requiring a heavier-than-Salpeter IMF as inferred if the central component is neglected. 
Likewise, allowing for a significant $M_{\rm cen}$ yields consistency with the NFW dark matter density profile, whereas fits neglecting the central mass required a
significantly  shallower profile.

To investigate further this apparent source of bias, we have performed simple simulations using idealised kinematic data tuned to the case of Abell 1201, which 
capture the effects of neglecting a compact central mass component. 
For each simulation, we use {\sc jam} generate a ``true'' velocity dispersion field from an input set of fixed model parameters 
($M_{\rm ap}$\,=\,34\,\mten, $\gamma$\,=\,1.0, $\Upsilon$\,=\,4.0, and $\beta_z$\,=\,0.2), and a range of central masses $M_{\rm cen}$ up to 3\,\mten. 
To derive the stellar mass component, we use the same MGE description of the luminosity density as employed in the main analysis. 
The kinematics are computed for the same spatial bins used for the real data. 
Subsequently, we fit these mock data with a two component (either NFW+Stars or gNFW+Stars) model, using the same machinery as applied to the real data. 
In doing so, we assign the bins the appropriate ``errors'' taken from the observed data, to ensure the correct relative weighting of the data points.
Figure~\ref{fig:bhmatters} shows the results of this analysis. 
The test confirms that {\it if} a compact central component is present, and is neglected in the fitting, this leads to a substantial overestimate of 
$\Upsilon$ (by $\sim$60 per cent for $M_{\rm cen}$\,=\,2\,\mten) and a severe underestimate of $\gamma$ (recovering $\gamma$\,=\,0.6 for $M_{\rm cen}$\,=\,2\,\mten),
if using the gNFW form for the halo. 
Note that the biases in $\Upsilon$ and $\gamma$ in the   NFW+Stars fits are found even when fitting essentially perfect simulated data, 
differing only in the dynamical influence of the central mass. In turn this suggests that the similar sensitivity of the {\it real} results to the 
inclusion of $M_{\rm cen}$ does not somehow arise from the high $\chi^2$ of the fit to the observed velocity field, 
or from mismatches in  spatial resolution, or other effects inherent to the observations. 

Returning now to the context of previous work on BCGs, recall that  \cite{2013ApJ...765...25N} fitted a gNFW+Stars model to 
seven clusters at $z$\,=\,0.2--0.3,  and found
evidence for {\it both} a shallower halo slope $\gamma$\,$\approx$\,0.5 {\it and} typical stellar mass-to-light ratios characteristic of the Salpeter IMF. 
These conclusions are strikingly similar to the results we obtain for Abell 1201, when $M_{\rm cen}$ is omitted from the dynamical model. 
The Newman et al. BCGs are more distant on average, and they argued that central black holes should
not be dynamically significant given the slit width and spatial resolution of their spectroscopy.
This is probably a reasonable assumption for black holes at the mean of the local scaling relations, but its validity
clearly depends on the maximum $M_{\rm BH}$ considered plausible.
It is outside the scope of this paper to explore fully whether the neglect of central black holes might have 
contributed to the conclusions reached by Newman et al. A meaningful study would need to account for important differences in the data available for their work
compared to ours. In particular, while our analysis benefits from the presence of a small-radius strong-lensing constraint in Abell 1201, 
Newman et al. were able to include weak-lensing shear profiles at $>$100\,kpc, as well as strong-lensing constraints out to 50--100\,kpc scales 
(a factor of ten larger than in Abell 1201), to anchor the outer dark matter profiles. The relative advantages afforded by these different configurations, 
and in particular their sensitivity to an additional central mass, are not trivial to establish.

A less obvious, but intriguing, comparison can be made with the results found by \cite{2016MNRAS.456..538Y} from orbit-based dynamical models of 
the compact elliptical galaxy NGC\,1281 in the Perseus cluster. At face value, 
this is a very different kind of galaxy than the massive and diffuse BCG in Abell 1201, but the authors report a qualitatively similar interplay between stars, dark matter and
black hole in their modelling. As in Abell 1201, the chain of constraints is that the outer kinematics require large dark matter contributions at intermediate radius, which drives
down the the stellar component, which in turn necessitates a massive central black hole to fit the central kinematics. In NGC\,1281, the best-fitting parameters
are arguably unreasonable (a much lighter-than-Kroupa IMF and an implausibly large halo-to-stellar mass ratio), and the authors were not able to account for the results 
within their modelling framework. While these findings should motivate caution in interpreting our results, we stress that in Abell 1201, by contrast,
the derived parameters {\it do} appear physically plausible, despite requiring a central mass component at the limit of current estimates for black holes.
 
\section{Conclusions}\label{sec:conc}

We have measured and analysed the spatially-resolved stellar kinematics for the strong-lensing BCG in Abell 1201. The large radial extent and 
high spatial resolution of our observations, combined with the unusual lensing configuration in this system, allow us to decouple three components
of the mass distribution in the cluster: stars, dark matter, and an additional central mass, which could be a super-massive black hole. 

Our models strongly favour the presence of a $>$\,10$^{10}$\,M$_\odot$ compact central mass, a result which is independently supported by the 
lensing analysis (Paper I). The best-fitting central mass of (2.5$\pm$0.2)\,\mten\ is an order of magnitude higher than expected from the local $M_{\rm BH}$--$\sigma$ relation, 
and comparable to the largest black hole masses measured to date, some of which also reside in BCGs. As an alternative to a black hole, the central mass could be
due to a strongly non-uniform stellar mass-to-light ratio, e.g. from a bottom-heavy IMF affecting only the innermost part of the galaxy.

Accounting explicitly for the central mass, we have highlighted the changes which follow in the other fitted components.
For an NFW halo, the dark matter contribution increases (from 36 per cent inside
the fiducial aperture to 54 per cent), and the stellar mass-to-light ratio is correspondingly reduced, from $\Upsilon$\,$\approx$\,6.7 to  $\Upsilon$\,$\approx$\,3.9, when the 
central mass is included. For comparison, an IMF similar to that of the Milky Way \citep{2001MNRAS.322..231K} predicts a value of $\Upsilon$\,$\approx$\,3.4, given the high
metallicity and probable early formation time of the BCG stellar population. Hence, including the central mass component in the model largely removes the need for a heavy IMF in this galaxy. 

Finally, considering a gNFW dark matter density profile with a free inner slope $\gamma$, we find once again that the treatment of a compact central mass is crucial. 
When fitting the lensing and kinematic data using only the halo and stellar components, the models yield $\gamma$\,$\approx$\,0.5, similar to the results found
by \cite{2013ApJ...765...25N}. By contrast, when we account for the presence of the central mass component, we infer $\gamma$\,=\,1.0\,$\pm$\,0.1, i.e.
no deviation from the standard NFW halo is required.

Future AO-assisted IFU data for the Abell 1201 BCG would add confidence to the detection of a very massive central black hole by better resolving its dynamical influence. Meanwhile, 
recently acquired 
{\it HST} observations with Wide Field Camera 3 
will improve the characterisation of the inner luminosity profile, and yield improved depth, resolution and lens-vs-source contrast.
A future paper will exploit these advances to present a refined lensing analysis of this system.

\section*{Acknowledgements} 
RJS, JRL \& ACE acknowledge support from the STFC through grants ST/L00075X/1 and ST/P000541/1. The datasets used for 
this paper are publicly available from the ESO Science Archive Facility. We thank Michele Cappellari and collaborators for releasing
and maintaining several software packages used in this paper.

\bibliographystyle{mnras}
\bibliography{rjs}

\bsp	
\label{lastpage}
\end{document}